# Inner Edge of Habitable Zones for Earth-sized Planets with Various Surface Water Distributions

**T. Kodama[1,2], H. Genda[3], R. O'ishi[2], A. Abe-Ouchi[2, 4], and Y. Abe[5, *]**

[1] Laboratoire d'astrophysique de Bordeaux, University of Bordeaux, Pessac, France.
[2] Center for Earth surface system dynamics, Atmosphere and Ocean Research Institute, The University of Tokyo, Chiba, Japan.
[3] Earth-Life Science Institute, Tokyo Institute of Technology, Tokyo, Japan.
[4] Japan Agency for Marine-Earth Science and Technology, Kanagawa, Japan.
[5] Department of Earth and Planetary Science, The University of Tokyo, Tokyo, Japan.
[*] Deceased.

Corresponding author: Takanori Kodama (takanori.kodama@u-bordeaux.fr)

**Key Points:**

- Climates for various surface water distributions are calculated via three-dimensional general circulation models.

- The insolation when a planet enters the runaway greenhouse state depends on the surface water distribution.

- A small difference in the amount of surface water forms a significant difference in climate between an aqua planet and a land planet

## Abstract

When planets receive insolation above a certain critical value called the runaway threshold, liquid surface water vaporizes completely, which forms the inner edge of the habitable zone. Because land planets can emit a large amount of radiation from the dry tropics, they have a higher runaway threshold than aqua planets do. Here we systematically investigated the runaway threshold for various surface water distributions using a three-dimensional dynamic atmosphere model. The runaway threshold for the meridionally uniform surface water distribution increases from the typical value for the aqua-planet regime ($\sim 130\%$ $S_0$) to one for the land-planet regime ($\sim 155\%$ $S_0$) as the dry surface area increases, where $S_0$ is the present Earth's insolation. Although this result is similar to the previous work considering zonally uniform surface water distributions, the runaway threshold for the land-planet regime is quite low compared to that of the previous work. This is because a part of the tropical atmosphere is always wet for the meridionally uniform case. We also considered the surface water distributions determined by the Earth's, Mars' and Venus' topographies. We found that their runaway thresholds are close to that for the meridionally uniform cases, and the amount of water at the boundary between an aqua- and land-planet regime is around 10% of the Earth's ocean. This clearly shows that the runaway threshold is not determined uniquely by the luminosity of the central star, but it has a wide range caused by the surface water distribution of the terrestrial water planet itself.

## Plain Language Summary

Some of the detected exoplanets are expected to be Earth-sized rocky planets, probably with liquid water on their surface due to an adequate orbital distance from the central star. The region where liquid water is stable on the planetary surface is called the habitable zone. Previous studies showed that the habitable zone for a planet with a small amount of water (called land planets) is wider than that of a planet with a large amount of water (aqua planets). By using a three-dimensional model for the atmospheric circulation, we investigated the inner edge of the habitable zone considering the effects of various surface water distributions. As the previous studies have done, we also recognized two climate regimes (aqua planet regime and land planet regime). The insolation that a planet receives at the inner edge of the habitable zone increases with an increase of the dry surface area. We considered the real landform patterns of Earth, Venus and Mars, and found that the boundary between the two climate regimes appears when the total amount of the surface water is around 10% of the Earth's ocean. Our results will give us a strong constraint on the amount of surface water in future observations using the significant difference between climates for an aqua planet and a land planet.

## 1 Introduction

We know many exoplanets that are considered to be terrestrial planets based on the relationship between planetary mass and radius [e.g., *Batalha et al.*, 2013]. Some of them are in the habitable zone of their host star. The habitable zone is defined as the region at such a distance from the host star that liquid water on the planetary surface remains stable [*Kasting et al.*, 1993; *Kopparapu et al.*, 2013; 2014]. Many studies have focused on determining the boundary of the habitable zone because planets within the habitable zone will be primary candidates for future observations to detect a biomarker in the search for life.

The boundaries of the habitable zone have been estimated using not only one-dimensional climate models but also three-dimensional general circulation models (GCM). Regarding the inner edge of the habitable zone, there are two definitions: one is the moist greenhouse limit and the other is the runaway greenhouse limit [*Kasting et al.*, 1993; *Kopparapu et al.*, 2013; 2014]. The difference between the moist greenhouse state and the runaway greenhouse state is whether the climate is climatologically stable or not. For the moist greenhouse climate, the climate is stable with a surface temperature of ~340 K, while the runaway greenhouse climate experiences uncontrolled warming and its surface temperature increases up to ~1600 K [*Goldblatt et al.,* 2013]. For the former one, it is defined by the lifetime of the ocean against an escape of water into space. When a planet with water on the planetary surface receives strong radiation from the host star, the mixing ratio of water vapor in the upper atmosphere increases and rapid water loss occurs [*Hunten*, 1973; *Walker*, 1977]. If this mixing ratio is above ~$3\times10^{-3}$, a planet will lose an amount of water equivalent to the Earth's ocean mass to space in ~4.5 Gyrs. This stage is called the moist greenhouse state. This threshold corresponds to the inner edge of the continuously habitable zone when we consider the long-term stability of liquid water on the planetary surface.

The latter one is defined by the balance between the net absorbed solar radiation and outgoing infrared radiation (called planetary radiation). The planetary radiation has a limit in the case of a wet atmosphere [*Ingersoll*, 1969; *Nakajima et al.*, 1992]. This limit is called the Simpson-Nakajima limit [*Goldblatt et al.*, 2013]. When a planet with liquid water on the planetary surface receives insolation above the Simpson-Nakajima limit from the host star, it is no longer able to maintain a thermal equilibrium state, and all of the liquid water on the planetary surface evaporates. This phenomenon is called the runaway greenhouse effect [e.g., *Komabayashi*, 1967; *Ingersoll*, 1969; *Kasting*, 1988; *Abe and Matsui*, 1988]. The threshold of the runaway greenhouse effect corresponds to the inner edge of the instantaneous habitable zone.

*Nakajima et al.* [1992] investigated the relationship between the temperature on the planetary surface and the outgoing infrared radiation using a one-dimensional radiative-convective equilibrium model. In a case with an atmosphere completely saturated by water vapor, they found that the outgoing infrared radiation that a water terrestrial planet can radiate has a limit of a particular value because the atmospheric pressure and temperature structures approach the saturated vapor pressure curve of water vapor as the surface temperature increases. The runaway threshold is estimated to be 282 W/m$^2$ (102% $S_0$) [*Goldblatt et al.*, 2013] and 288 W/m$^2$ (104% $S_0$) [*Kopparapu et al.*, 2013], depending on their one-dimensional climate models, where $S_0$ is the insolation for a G-type star, like our Sun, at present Earth's orbit.

The values for the runaway threshold estimated by three-dimensional general circulation models tend to be higher than that estimated by one-dimensional climate models. This is because an unsaturated region of water vapor appears in 3-D models, which is formed by the descending flow of the Hadley circulation [*Abe et al.*, 2011; *Leconte et al.*, 2013b; *Wolf and Toon*, 2014; 2015]. *Leconte et al.* [2013b] estimated the runaway threshold to be 110% $S_0$ by using the LMD (Laboratoire de Météorologie Dynamique) generic GCM. *Wolf and Toon* [2015] showed that the thermal equilibrium state is maintained up to 121% $S_0$ of the insolation using the CAM4 (Community Atmosphere Model version 4). *Way et al.*, [2018] showed a stable climate maintains up to at least 120% $S_0$ for the Earth by using ROCKE-3D.

These estimations of the runaway threshold in previous studies conventionally assumed a water planet with a large amount of water on the planetary surface, which implies that its surface is globally covered with liquid water. In this study, we call such a planet "an aqua planet". However, it becomes clear that the amount of water that a planet has strongly affects the planetary climate [*Abe et al*., 2005; *Abe et al*., 2011; *Abbot et al*., 2012; *Leconte et al*., 2013a; *Kodama et al*., 2018]. *Abe et al*. [2011] investigated the climate of an idealized planet with a small amount of water on its surface (called a land planet), using CCSR/NIES AGCM5.4g (Center for Climate System Research/National Institute for Environmental Studies Atmospheric General Circulation Model 5.4g). The most important character of the climate for this planet is that liquid water distribution is localized near the poles, due to the precipitation and the evaporation being balanced locally, leading to very low relative humidity in the planetary atmosphere in dry regions. As a result, a land planet can radiate a larger outgoing infrared radiation from a dry equator region than an aqua planet can. Therefore, a land planet can maintain liquid water on its surface up to 170% $S_0$. This value is quite a lot larger than the Simpson-Nakajima limit.

*Kodama et al*. [2018] investigated the effect of the surface water distribution on the runaway threshold. They considered a zonally uniform distribution of surface water introducing a new conceptual parameter called the water flow limit. The planetary surface is assumed to always be wet in the higher latitude region than the latitude of the water flow limit. They performed GCM calculations for this hypothetical planet. They found that a planet behaves as an aqua planet and the runaway threshold is a constant value (about 130% $S_0$) when a planet has a wet surface region extending from both poles to latitudes lower than about 30°. In this case, the water vapor is transported to the equator region by the Hadley circulation, which makes the atmosphere globally wet. Therefore, the outgoing infrared radiation is limited. As the lowest latitude of the wet surface region approaches the poles, the runaway threshold varies continuously from ~130 $S_0$ to ~180% $S_0$, which indicates that such planets behave like land planets. Thus, whether a water planet behaves as an aqua planet or a land planet is determined by the relationship between the lowest latitude of the wet surface region and the location of the Hadley circulation.

However, *Kodama et al*. [2018] focused only on the zonally uniform surface water distribution. Thus, it is necessary to carefully consider the relationship between the runaway threshold and the water distribution on more realistic planets with oceans. The surface water distribution is determined by the planetary topography. First, in this study, we investigate the dependence of the runaway threshold on the meridionally uniform surface water distribution using the same methods as *Kodama et al*. [2018]. Second, we consider various surface water distributions with different amounts of water using the planetary topography of terrestrial planets in our solar system, and estimate the runaway threshold in cases of created surface water distributions. Finally, taking these results, we aim to comprehensively understand the effect that the water distribution on the planetary surface has on the runaway threshold.

In Section 2, we shortly describe our GCM. In Section 3, we show some typical results for the meridional water distribution and the dependence of the runaway threshold on the surface water distribution. We also show results for a planet with 2-D surface water distribution which is determined by the planetary topographies in Section 4. Then, in Section 5, we discuss the amount of water at the boundary between an aqua planet and a land planet and the effect of oceanic albedo on the runaway threshold. In Section 6, we summarize our findings.

## 2 GCM description

To understand the dependence of the runaway threshold on the surface water distribution, we use three-dimensional GCM, CCSR/NIES AGCM5.4g, which is the same code as that in *Kodama et al.* [2018]. Although we set the same setting as that of *Kodama et al.* [2018] except for the initial surface water distribution, here we describe the essential points in this study. For more details, see *Kodama et al.* [2018] and *Abe et al.*, [2011].

This GCM was developed for the modeling of the present Earth's climate [*Numaguchi*, 1999] and was applied to the paleoclimate of Earth [*Abe-Ouchi et al.*, 2013]. The resolution of GCM is about 5.6º in latitude and longitude, and the number of vertical layers is 20 layers with the sigma coordinate. The fundamental equations of dynamical processes are primitive equations [*Haltiner and William*, 1980]. The large-scale condensation is determined by the scheme of prognostic cloud water based on *Le Treut and Li* [1991]. The scheme of cumulus precipitation is based on a simplified Arakawa-Schubert scheme [*Arakawa and Schubert*, 1974; *Moorthi and Suarez*, 1992]. The two-stream $k$-distribution scheme is adopted for the scheme of radiative transfer, which has 18 wavenumber channels, which are divided by 50, 250, 400, 770, 990, 1100, 1400, 2000, 2500, 4000, 14500, 31500, 34500, 36000, 43000, 46000 and 50000 cm$^{-1}$, with several sub-channels. The total number of channels is 37 [*Nakajima and Tanaka*, 1986]. We used the bucket model to estimate soil water content [*Manabe*, 1969]. The soil has the ability to contain liquid water up to the field capacity $W_{g,max}$. We set it to be 1000 m to avoid the surface runoff. The soil moisture $W_g$ is predicted as a net contribution of rainfall, evaporation and snowmelt. To estimate the evaporation efficiency, we introduce the critical soil moisture $W_{g, crit}$, which is the threshold of soil moisture for runoff and is set to be 10 cm. When $W_g$ is greater than 10 cm, the surface is assumed to be entirely wet and the evaporation efficiency $\beta$ is unity. On the other hand, when $W_g$ is less than 10 cm, $\beta$ is the ratio between soil moisture and critical soil moisture $W_g/W_{g, crit}$. When the surface is completely dry, $\beta$=0. Here, the efficiency of evaporation from the soil surface is determined as a function of the soil moisture and potential evaporation, which is the hypothetical evaporation efficiency from a completely wet surface. The potential evaporation from the surface is estimated by a bulk formula. The amount of snow is predicted by the net contribution between snowfall and snowmelt. The surface albedo without snow is 0.3, which is a typical value in the desert. We did not take into account the effect of decreasing albedo for soil saturation even if water pool region. This gave us the lower limit for the runaway threshold. We built idealized water planets from Earth, where the radiative effect of ozone, vegetation and the energy transportation of oceans are removed. These idealized water planets are the same as what *Abe et al.* [2011] and *Kodama et al.* [2018] used. We assumed that a fixed 1 bar of N$_2$ as the atmospheric composition which does not include CO$_2$, planetary size (surface gravity) and orbital period which were unchanged from the present Earth's values; the planetary orbit is circular; and the planetary obliquity is zero to avoid the effects of seasonal changes on the insolation.

## 3 Cases for the meridional water distribution

### 3.1 Method

The procedure of estimation of the runaway threshold is the same as that of *Kodama et al.* [2018] except for considering the meridionally uniform surface water distribution. The planetary surface is divided into 64 grids in longitude in our GCM, and some of them are set as the water pool region, which is always wet, from the North Pole to the South Pole. As the initial condition for our simulations, we ran a GCM simulation keeping the land surface saturated for 10 years. Then, we put water of 100 m depth in the water pool region and 10 cm depth in the regions outside of the water pool region. The evaporation efficiency is globally unity at the initial condition. If we put different depths of water in the region outside the water pool region, the surface water distribution will reach a stable condition because the distribution of water on the surface is determined by the precipitation and evaporation. We perform GCM calculation with the insolation of the present Earth for 10 years until a stable climate is achieved. After that, we reset the thickness of water: 100 m depth in the water pool region, 10 cm depth in the region where $\beta$=1 and the thickness in the rest are unchanged. We introduced this resetting procedure to save calculation time to get the thermal equilibrium state. For the water pool region, the change in the depth of water in the water pool region does not affect the results as long as the depth of water remains above 10 cm (i.e., $\beta$=1) during GCM calculation. The depth of 100 m in the water pool region is enough to keep $\beta$=1 for the next GCM calculation. For regions outside the water pool region where $\beta$=1 (i.e., more than 10 cm), we also reset the depth of water to 10 cm. If the precipitation minus evaporation in these regions goes negative for the next GCM calculation, the surface should become dry even if its depth of water is more than 10 cm in a previous step. On the other hand, if the precipitation minus evaporation becomes positive in these regions, the depth of water in these regions should increase. Then, we perform GCM calculation with increased insolation from the atmosphere condition on one step before with resetting surface water distribution until the climate reaches a thermal equilibrium where the radiation budget (the net absorbed solar radiation minus the outgoing planetary radiation) is balanced within about 1 W/m$^2$. We repeat these steps until the climate cannot reach the equilibrium state and the calculation gets unstable, and determine the runaway threshold as the highest insolation for keeping the thermal equilibrium state (see Figure S1).

We consider two cases for the water distribution on the surface: the meridionally equally dispersed case and the meridionally concentrated case. Hereafter, we call these two cases "the dispersed case" and "the concentrated case", respectively. We investigate the runaway threshold in the cases of 1-9 grids of the water pool region. Figure 1 shows schematic diagrams of the surface water distribution with 6 grids of the water pool region for the dispersed and concentrated case. Both cases have the same total area for the water pool region.

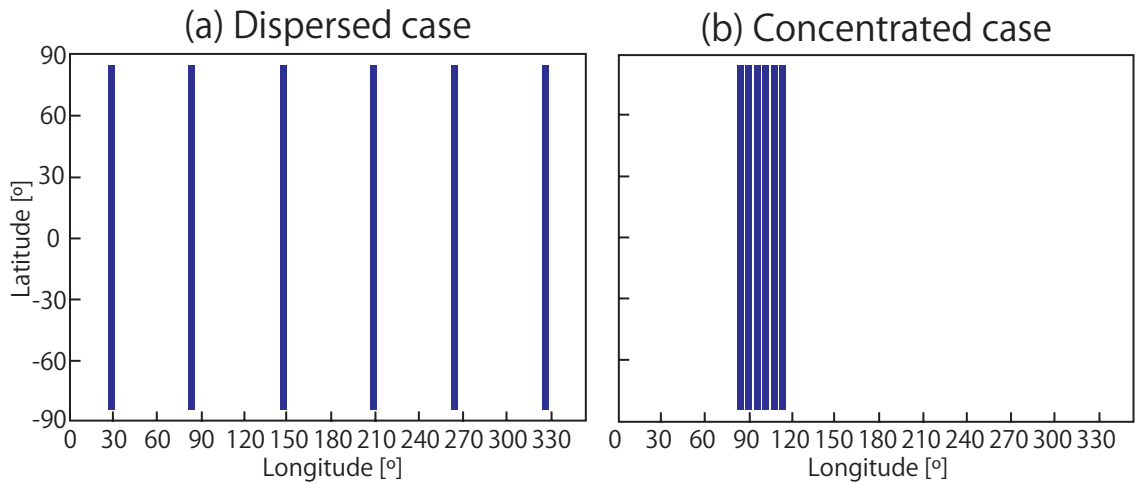

Figure 1. Schematic diagrams of surface water distribution in cases for (a) the meridionally equally dispersed case and (b) the meridionally concentrated case for 6 grids of the water pool region.

## 3.2    Runaway threshold

Figure 2 shows time-series plots of the bottom-of-atmosphere global mean temperature and the top-of-atmosphere energy budget (the absorbed shortwave radiation minus the outgoing longwave radiation) for the dispersed case and the concentrated case. The number of grids of the water pool region in both cases is 6 grids. Both cases show that an imbalance between the net absorbed solar radiation and outgoing infrared radiation occurs for runaway cases and the bottom-of-atmosphere global mean temperature increases when planets receive strong insolation from the central star, which means the planetary climate lapses into the runaway state. In the very last moment of the simulations, the net radiative budget goes to negative due to a temporal increase in the outgoing longwave radiation caused by the decrease of the relative humidity around the tropic on our calculation. Nevertheless, the global mean temperature increases. Since the climate is not in equilibrium state during the runaway greenhouse state, this temperature increase should be a temporal phenomenon. On the other hand, when the planet receives slightly weaker insolation than that of the runaway case, an imbalance of the radiation budget does not occur, and the planetary climate maintains a state of thermal equilibrium. In these stable cases, the water mixing ratio in the upper atmosphere is still quite low ($\sim 10^{-9}$), which is well below the water mixing ratio ($\sim 10^{-3}$) at the moist greenhouse threshold defined by the 1-D climate model [*Kasting* 1988]. The moist greenhouse atmosphere is thought to lead to a significant water loss. Our results are also consistent with *Abe et al.* [2011]. These cases do not easily enter the moist greenhouse state, although the resolution of this GCM is not enough to precisely investigate the mixing ratio of water vapor in the upper atmosphere. Whether the moist greenhouse state occurs or not is still remaining issue [*Kasting et al.*, 2015]. *Leconte et al.* [2013b] found the cold stratosphere which leads to a low water mixing ratio. On the other hand, *Wolf and Toon* [2015] found much warmer stratosphere with a higher water mixing ratio at high surface temperature. Thus, more accurate simulations are required to investigate the moist greenhouse atmosphere. Here, we focus on the onset of the runaway greenhouse state. We define the runaway threshold as the upper limit insolation below which the planetary climate can maintain a thermal equilibrium state. In these cases, the runaway thresholds are 129% $S_0$ and 142% $S_0$ for the dispersed case and the concentrated case with 6 grids of the water pool regions, respectively. The runaway thresholds are different even for the same total area of the water pool region.

Figure 3 shows climate variables at the runaway threshold (129% $S_0$) for the dispersed case with 6 grids of the water pool regions. Since we put the water pool region over equal intervals (Figure 3a), regions with $\beta=1$ have similar intervals between locations (Figure 3b). In these regions, liquid water evaporates efficiently (Figure 3d), water vapor is transported to the atmosphere and the precipitable amount of water becomes larger near the equator (Figure 3c). As a consequence, the outgoing long-wave radiation near the equator cannot exceed the Simpson-Nakajima limit ($\sim$300 W/m$^2$) because the atmosphere is wet (Figure 3e).

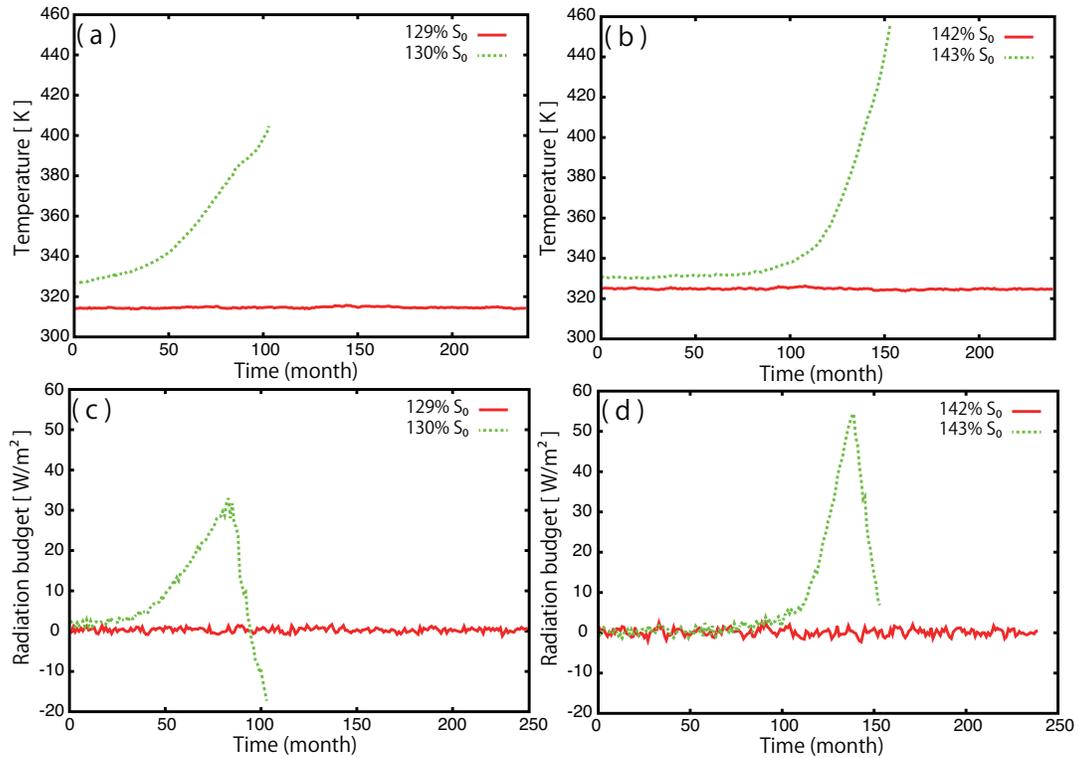

Figure 2. Time series plots of the temperature at the bottom of the atmosphere (a and b) and the radiation budget at the top of the atmosphere (c and d) for meridionally equally dispersed case (left column) and the meridionally concentrated case (right column) for 6 grids of the water pool region, respectively. The values of the incident solar radiation are shown in each panel in the unit of % $S_0$.

Figure 4 also shows climate variables at the same insolation (129% $S_0$) as that of Figure 3, but for the concentrated case. The number of grids for the water pool region is 6 grids in both Figure 3 and 4; that is, both cases have the same area of water pool region. In the water pool region, the evaporation efficiency $\beta$ is unity and efficient evaporation occurs (Figure 4b and d) as shown in Figures 3b and d. However, the precipitable amount of water for the concentrated case (Figure 4c) is less than that for the dispersed case (Figure 3c). Thus, the region where the outgoing long-wave radiation is limited by the wet atmosphere is narrower than that for the dispersed case (Figure 3e and 4e). As a result, the concentrated case does not reach the runaway threshold at 129% $S_0$ and can sustain a thermal equilibrium state against stronger insolation.

We repeat increasing the insolation by 1% $S_0$ for this concentrated case with 6 grids of the water pool region until an imbalance of the radiation budget occurs. Figure 5 shows the climate variables for this case at the runaway threshold (142% $S_0$). More efficient evaporation occurs in the water pool region (Figure 5d). The amount of precipitable water gets larger (Figure 5c) at almost all longitudes near the equator. Therefore, the planetary climate reaches the state of the runaway threshold because of the limit in the outgoing long-wave radiation for a moist atmosphere.

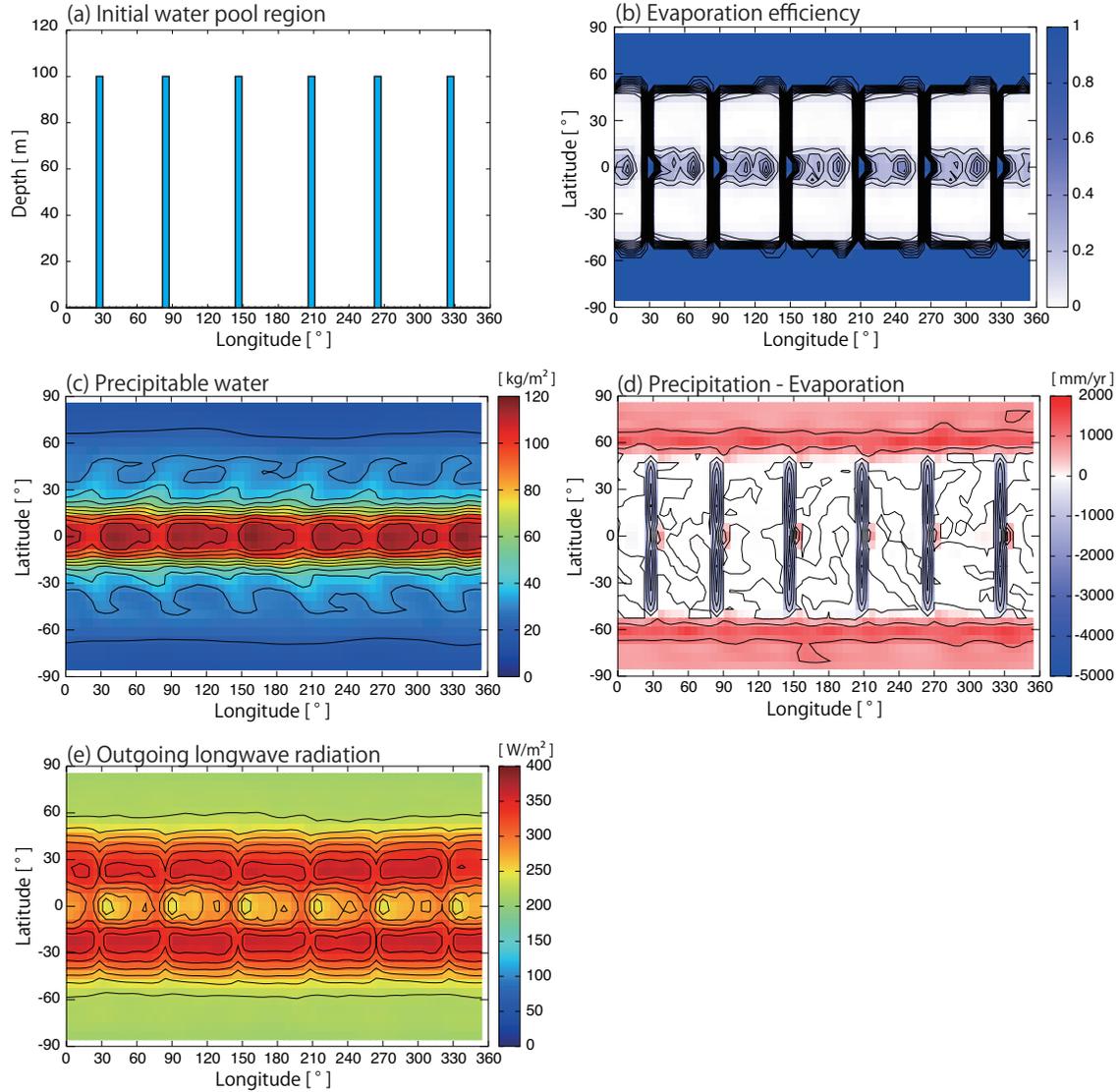

Figure 3. Climate variables for the meridionally equally dispersed case with 6 grids of the water pool region for 129% $S_0$, which corresponds to the runaway threshold. Each panel shows (a) the initial depth of water; (b) the evaporation efficiency $\beta$ expressing the ground wetness; (c) the precipitable water; (d) the precipitation flux minus the evaporation flux; (e) the outgoing long-wave radiation.

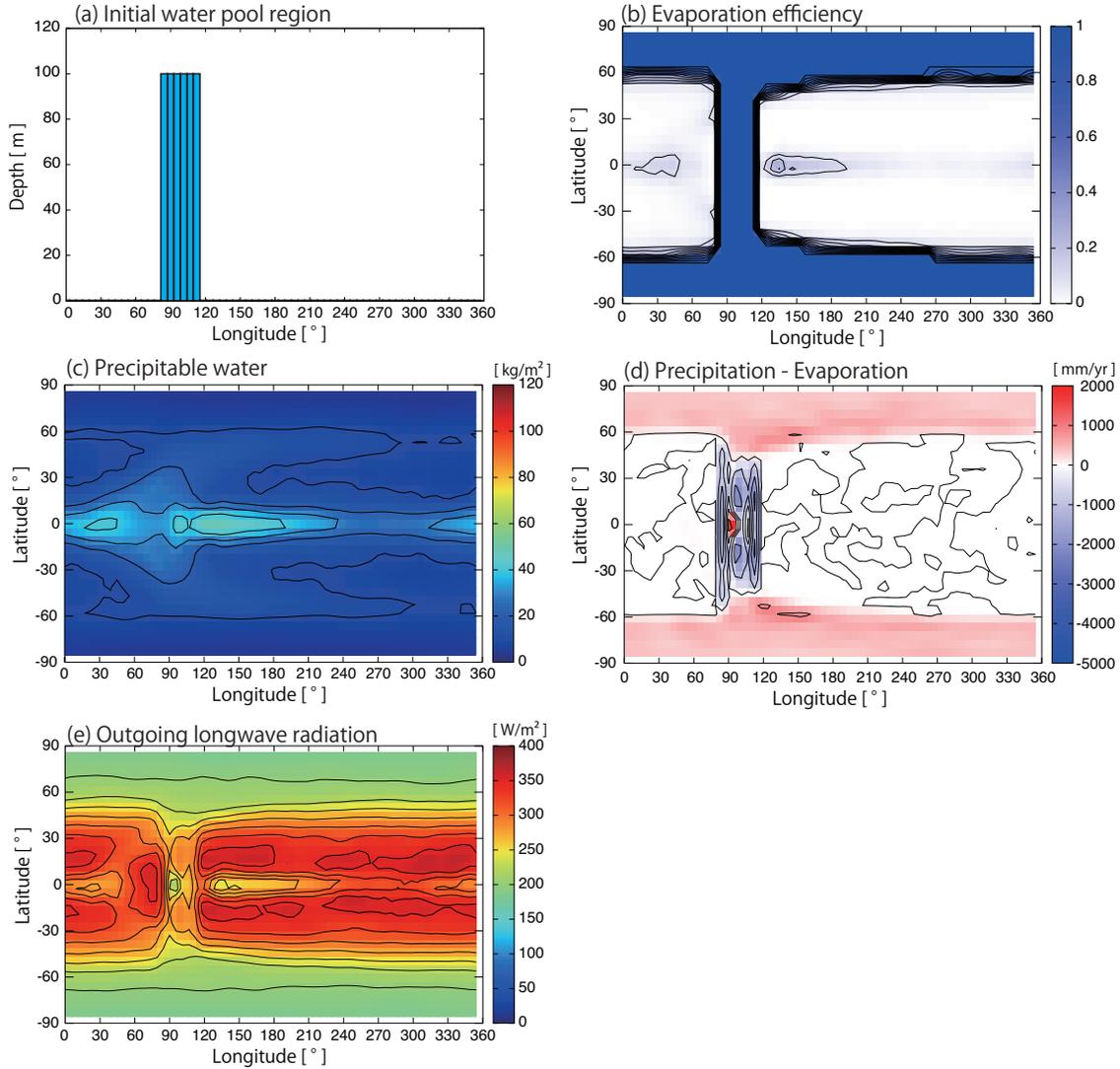

Figure 4. As in Figure 3, climate variables for the meridionally concentrated case with 6 grids of the water pool region for 129% $S_0$. Climate in this case is still stable.

The insolation at the runaway threshold for this case (142% $S_0$) is larger than that for the meridionally equally dispersed case (129% $S_0$).

The difference in the insolation of the runaway threshold for cases with the same area of water pool region, but different surface water distribution comes from the distribution of the precipitable water in the atmosphere (see Figure 3c and 5c). The dispersed cases can efficiently wet the atmosphere near the equator because supply of water vapor from separated water pool regions keep the air parcel wet over the equator. On the other hand, for the concentrated case, it is necessary to evaporate a large amount of water vapor from the concentrated source of water vapor to reach the runaway threshold because the air parcel tends to become dry due to rainout until it circulates over the equator. Thus, larger insolation is required to cover the area near the equator with a large amount of the precipitable water. Additionally, the strength of the atmospheric circulation is also important to create the difference between two climates because it determines the distribution of the precipitable water in the atmosphere.

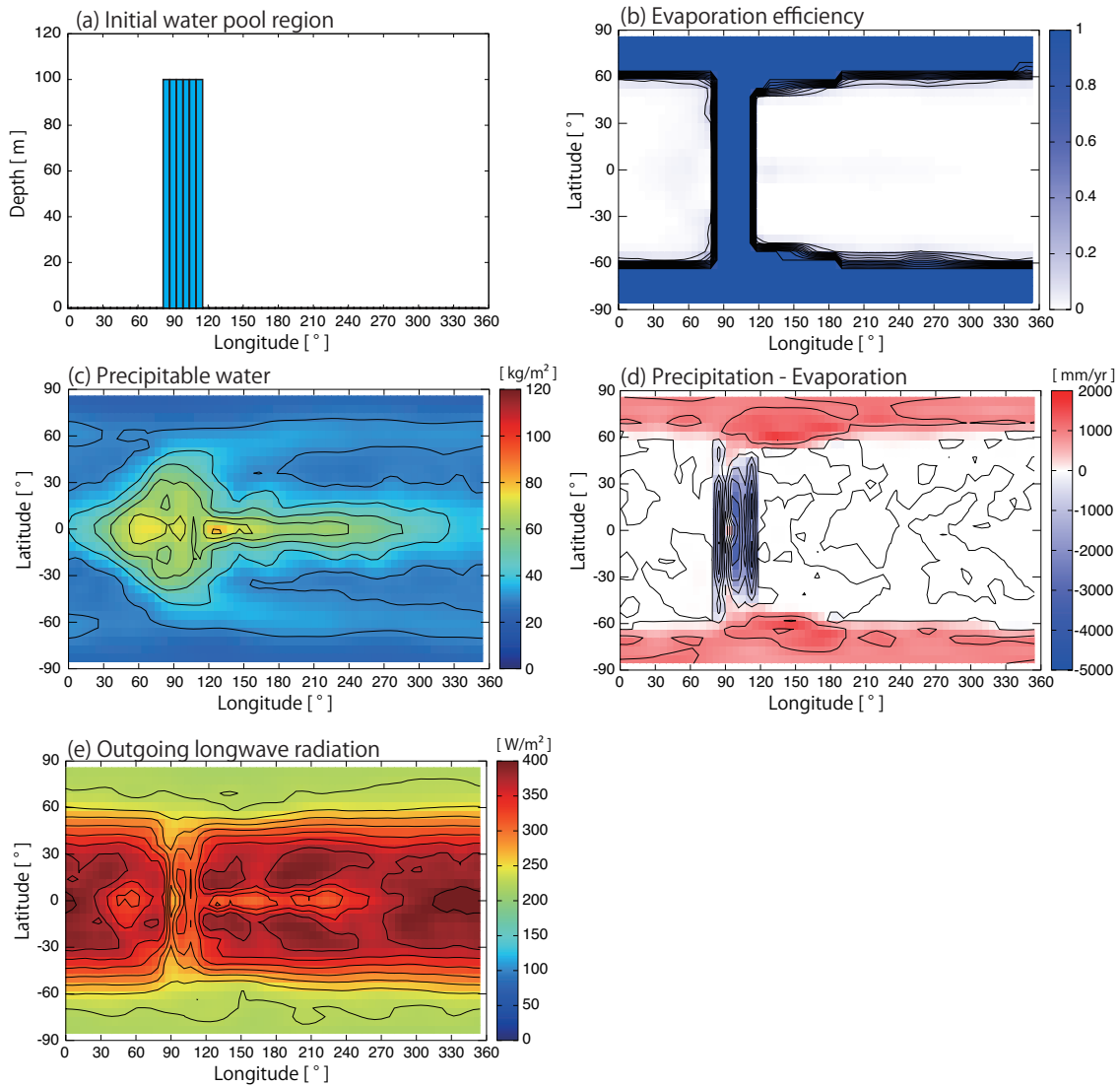

Figure 5. As in Figure 3, climate variables for the meridionally concentrated case with 6 grids of the water pool region, but for 142% $S_0$, which corresponds to the runaway threshold.

We showed the climate variables for the cases with the same area of the water poor region (6 grids) but different surface distribution (Figure 3-5). Next, we show the results for the cases with a different area of the water pool region ranging from 1 grid to 9 grids. Figure 6 shows the calculated runaway threshold for the dispersed and concentrated cases as a function of the number of water pool grids. For the same total area of the water pool region, the concentrated case would give the upper limit for the runaway threshold due to the least efficient water transportation to the atmosphere, and the dispersed case would give the lower limit for the runaway threshold due to the most efficient water transportation to the atmosphere. Thus, we can expect that the runaway threshold for any longitudinal surface water distribution should be located in between the results for both cases. The difference in the runaway threshold between dispersed and concentrated cases increases with the increase in the number of water pool grids. The difference reaches up to 13% $S_0$ at 6 grids of the water pool region (129% $S_0$ for the dispersed case and 142% $S_0$ for the concentrated case). After reaching the maximum, the difference remains almost constant, and the insolation of the runaway threshold gets close to the value for the aqua planet ($\sim$130% $S_0$) at 9 grids of the water pool region for the dispersed case. It is clear that more than 9 grids are needed for the concentrated case to behave as an aqua planet.

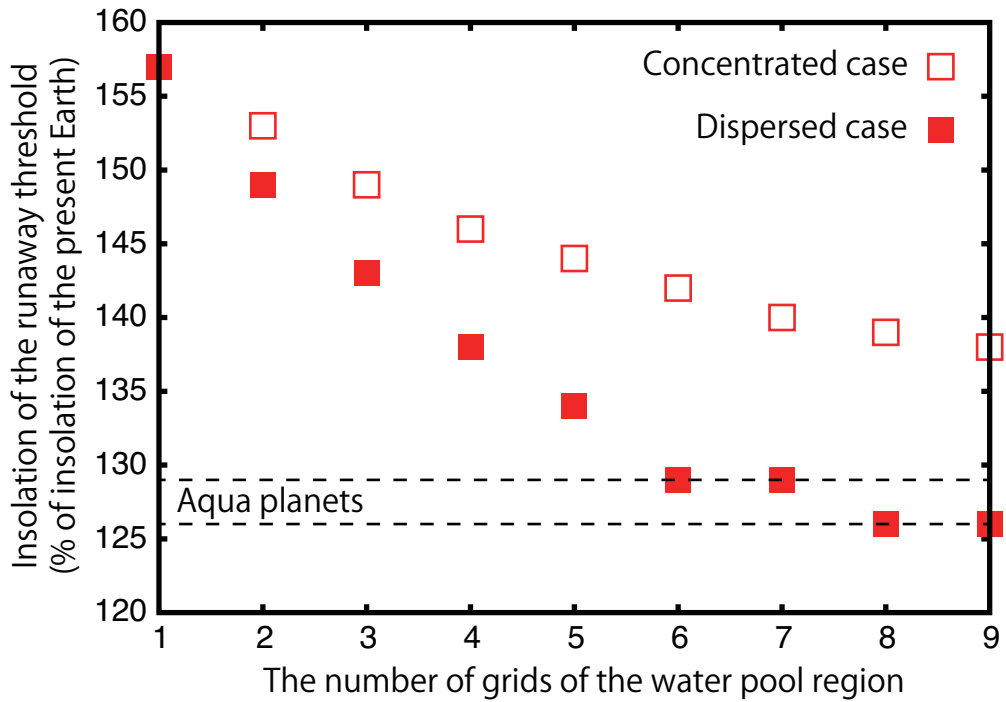

Figure 6. The runaway thresholds for the dispersed (filled square) and concentrated (opened square) cases as a function of the number of grids of water pool regions. The runaway threshold decreases with the increase in the number of grids of the water pool region. The runaway threshold for the dispersed case gets close to that of an aqua planet for more than 6 grids of water pool regions. The dashed lines describe the runaway threshold for aqua planets which also depend on the water distribution [*Kodama et al.,* 2018].

Next, we consider the dependence of the runaway threshold on a land fraction. Here we define the land fraction as the ratio of dry surface area divided by the total area. The water pool region is always wet, but the other regions are dry ($\beta<1$) or wet ($\beta=1$), which is predicted by the GCM calculation. We defined the regions with $\beta<1$ as the land. We calculated the area of the dry regions where $\beta<1$ to estimate the land fraction. Figure 7 shows the resultant runaway threshold as a function of the land fraction. The runaway thresholds for the dispersed and concentrated cases are similar for the same land fraction, although those are very different for the same number of water pool grids (Figure 6). The runaway threshold is determined by how wide the region with a large amount of water vapor in the atmosphere is. Thus, even if the area of the water pool region is the same, the evaporation efficiency is unity over a wider area for the dispersed case than for the concentrated case. In other words, the state of the climate for the dispersed case is close to that of an aqua planet because the planetary surface wets a wide region via precipitation.

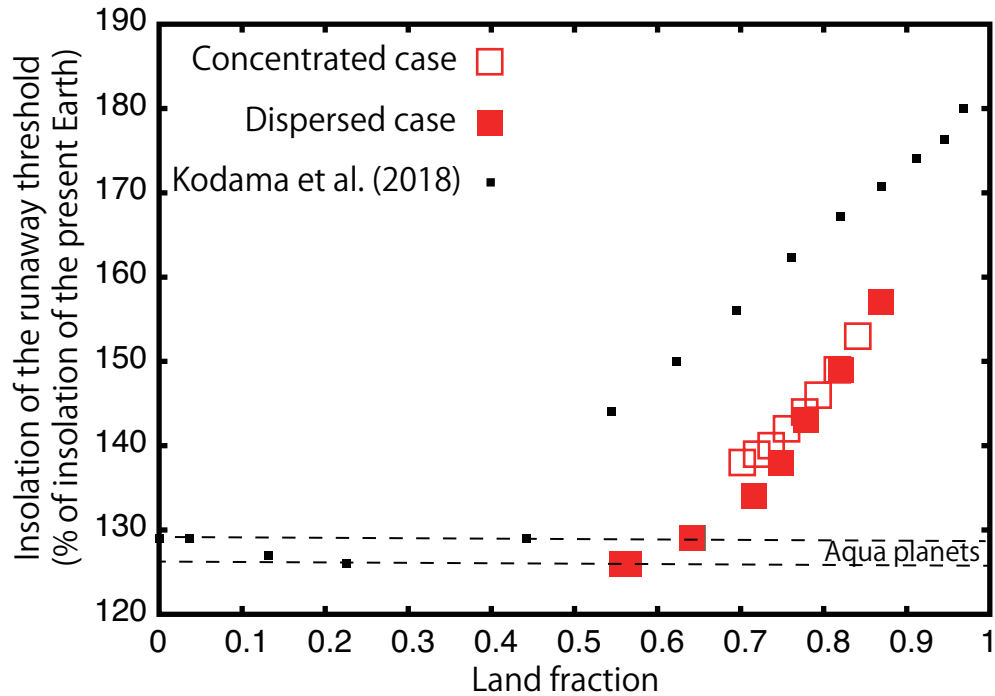

Figure 7. The runaway thresholds for concentrated and dispersed cases as a function of the land fraction. The results obtained in *Kodama et al.* [2018] (black square) are also shown. The runaway thresholds for the meridionally uniform surface water distribution is quite a lot lower than that for the zonally uniform case. The dashed lines describe the runaway threshold for aqua planets which also depend on the water distribution [*Kodama et al.,* 2018].

Figure 7 also shows the results obtained in *Kodama et al.* [2018]. They assumed a zonally uniform distribution of the surface water. Their setting of the initial condition of the surface water distribution wets the atmosphere near the equator less compared with our setting in this study, because the transportation of water vapor to the atmosphere efficiently occurs near the equator in this setting. Therefore, the setting in *Kodama et al.* [2018] gives the maximum insolation of the runaway threshold for a given land fraction and the setting in this study gives the minimum insolation of the runaway threshold. These results for two extreme cases give the width of the insolation of the runaway threshold caused by the surface water distribution. This clearly shows that the inner edge of the habitable zone is not determined uniquely by the luminosity of the central star, but it has a wide range caused by the surface water distribution of the terrestrial water planet itself.

## 4 Effects of planetary topographies for Earth, Mars and Venus

In the previous section, we investigated the runaway threshold for hypothetical and ideal surface water distributions. The water distribution on the real planets, which strongly depends on the strength of the water transportation on the surface, is governed by the amount of water and the planetary topography. Thus, given a specific planetary topography, the distribution of the water pool region would be determined by the amount of water on the planetary surface. In this section, we consider the planetary topography of the terrestrial planets in our solar system. Considering various water amounts on the planetary surface, we estimate the runaway threshold for the more realistic two-dimensional water distribution.

### 4.1 Topography for the terrestrial planets

We represent the topography of a terrestrial planet by spherical harmonics, based on *Wiezorek* [2006] and *Hirt et al.* [2012]. The planetary topography can be written with the corresponding expansion coefficients as below:

$$Topo(\theta, \phi) = \sum_{l=0}^{\infty} \sum_{m=0}^{l} (A_l^m \cos m\theta + B_l^m \sin m\theta) \overline{P_l^m}(\sin \phi), \quad (1)$$

where $l$ is a degree, $m$ is an order, $\theta$ is the longitude and $\phi$ is the latitude. The values of $\theta$ and $\phi$ have the range of $0 \le \theta \le 2\pi$ and $-\pi/2 \le \phi \le \pi/2$, respectively. $\overline{P_l^m}(\sin \phi)$ are normalized Legendre functions. These are expressed with Legendre functions $P_l^m(x)$ as

$$\overline{P_l^m}(x) = (-1)^{-m} \sqrt{(2 - \delta_{0,m})(2l+1)\frac{(l-m)!}{(l+m)!}} P_l^m(x), \qquad (2)$$

where

$$P_l^m(x) = (-1)^{-m}(1-x^2)^{m/2} \frac{d^m}{dx^m} P_l(x), \qquad (3)$$

and

$$P_l(x) = \frac{1}{2^l l!} \frac{d^l}{dx^l} (x^2 - 1)^l. \quad (4)$$

In particular, when $l = m$, then

$$P_m^m(x) = (2m-1)!! (1-x^2)^{m/2}, \quad (5)$$

$$P_{m+1}^m(x) = (2m+1)!! x(1-x^2)^{m/2}. \qquad (6)$$

All Legendre functions with any degree and order are calculated by the following recurrence formula with (5) and (6),

$$(l-m+1)P_{l+1}^m(x) - (2l+1)xP_l^m(x) + (l+m)P_{l-1}^m(x) = 0. \quad (7)$$

By summing up all normalized Legendre functions with each degree and order with expansion coefficients for the Earth, Mars and Venus, we obtain the topography of these terrestrial planets, as shown in Figure 8. Expansion coefficients we used are shown in the supporting information. Figure 8 shows the planetary topography of Earth, Mars and Venus with $l = m = 84$. We use those topography data with this resolution, which is high enough to be the spatial resolution of our GCM calculation.

### 4.2 Water pool regions as a function of water amount

After the representation of the planetary topography, we evaluate the distribution of surface water as a function of the surface water amount. In general, the transportation of water vapor from the equator to the pole is caused by atmospheric circulation and the surface water is transported from the pole to the equator. Thus, we pour water into a given planetary topography from the poles. In *Kodama et al.* [2018], they introduced the water flow limit as the lowest latitude that the surface water reaches. In this study, we follow this concept to determine the distribution of the surface water. When the water flow limit is given, we can estimate the amount of water by summing up the capacity of every depression that is located at a higher latitude than the water flow limit. Thus, we can get the relationship between the amount of water and the latitude of the water flow limit, and the distribution of surface water. When the water flow limit reaches the equator, we find the lowest altitude and the second lowest altitude, and the water was poured out until the lowest altitude reached the second lowest altitude.

We create seven water distributions with different amounts of water on the planetary surface ranging from $1 \times 10^{16}$ m$^3$ (~0.01 $V_{OEC}$) to $2.74 \times 10^{18}$ m$^3$ (~2 $V_{OCE}$) for the topography of Earth, Mars and Venus, where $V_{OCE}$ is the amount of the present Earth's ocean ($1.37 \times 10^{18}$ m$^3$). The 2-D surface water distributions with the different amounts of water are shown in the supporting information (Figure S1-S3). Then, we change the resolution of these water distributions to that of our GCM. In this study, we set all planetary radii to the Earth's value even when using Mars' and Venus' topography. We do not include the information of altitude of the planetary topography in GCM calculations. Although the planetary surface is globally flat in our simulations, the water pool region is always set to be wet and the other regions are calculated by using the GCM model. Because our purpose in this paper is to investigate the effect of the surface water distribution on the runaway threshold, we adopt this setting to avoid the physical phenomena caused by the differences in height on the planetary topography and planetary size.

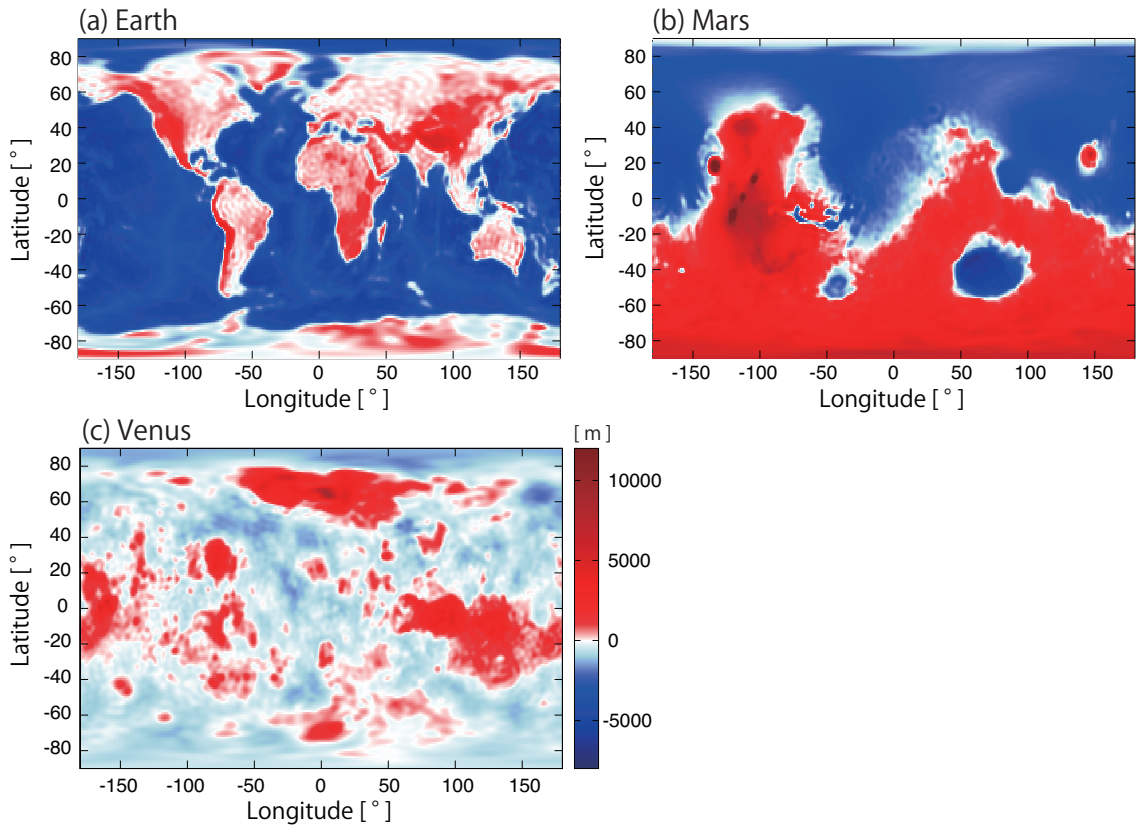

Figure 8. Planetary topographies created by using the spherical harmonics for (a) Earth case, (b) Mars case and (c) Venus case.

### 4.3 Runaway threshold

We determine the runaway threshold for the planets with a 2-D surface water distribution in the same procedure as was used in Section 3.2. Figure 9 shows the relationship between the runaway threshold and the land fraction for various amounts of water (namely various water distributions) using the terrestrial topographies in addition to the results for ideal water distributions shown in Figure 6. Overall, the results are divided into two modes around a land fraction of 0.4. In the case of a land fraction less than 0.4—in other words, a planet with a large amount of water on its surface—the insolation at the runaway threshold is close to that of an aqua planet and will be a constant value (~130% $S_0$). On the other hand, in the case of a land fraction larger than 0.4 where a planet has a small amount of water on its surface, the insolation at the runaway threshold is located between that for the zonally localized surface water distribution and the meridionally concentrated/dispersed surface water distribution. When a planet has a considerably small amount of water on its surface, the insolation at the runaway threshold is also close to the results in *Kodama et al.* [2018] because the distribution of the surface water is close to that for the zonally localized surface water distribution.

As discussed in Section 3.2, the insolation at the runaway thresholds for the zonally uniform surface water distributions corresponds to the upper limit of the insolation of the runaway threshold. On the other hand, these for the meridionally uniform surface water distributions correspond with the lower limit. The insolation at the runaway threshold is governed by the distribution of water vapor in the low-latitude regions where the atmosphere is wet enough to become opaque and radiate the planetary radiation of the Simpson-Nakajima limit. Our results for the 2-D water distribution using real topographies of terrestrial planets in our solar system have more or less depressions around the equator where the water is stored, and the evaporation efficiency is unity, which makes the atmosphere wet.

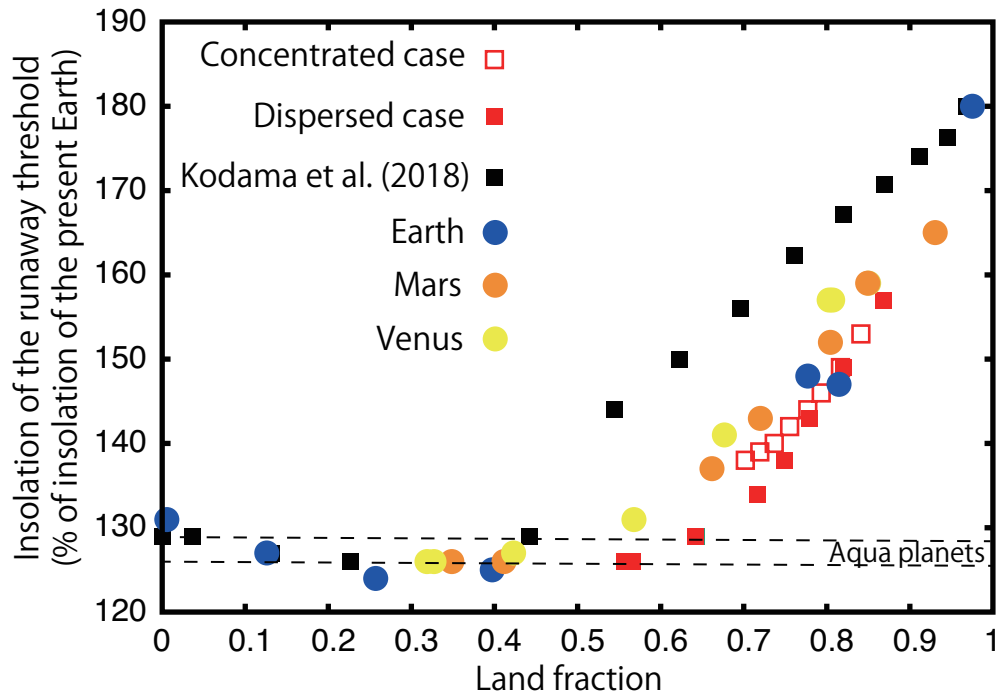

Figure 9. The runaway thresholds for various surface water distributions as a function of the land fraction. The runaway thresholds for water distributions determined by the Earth's, Mars' and Venus' topographies are located between those for the zonally and meridionally uniform surface water distributions. The dashed lines describe the runaway threshold for aqua planets which also depend on the water distribution [*Kodama et al.*, 2018].

### 4.4 Boundary between aqua and land planet's climates at the runaway threshold

Important differences between the aqua planet's climate and the land planet's climate at the runaway threshold are the evaporation efficiency and the distribution of the cloud mass mixing ratio. For an aqua planet, the evaporation efficiency is high and clouds form around the equator. The outgoing long-wave radiation is limited due to high relative humidity in the atmosphere. On the other hand, a land planet has a lower evaporation efficiency than that of an aqua planet and few clouds form around the equator. The outgoing long-wave radiation that a land planet can radiate is also larger than that of an aqua planet because of a dry atmosphere.

Figure 10 shows the radiation fluxes (the absorbed solar radiation and the outgoing long-wave radiation), the evaporation efficiency and the distribution of the cloud mass mixing ratio for both an aqua planet and a land planet at the runaway threshold to confirm the boundary of climates.

As an example of an aqua planet's climate, we show the results for the Earth's case with $1 \times 10^{17}$ m$^3$ of water, which corresponds to a land fraction of 0.40 (the left column in Figure 10). This climate is recognized as an aqua planet's climate by these variables in Figure 10. As an example of a land planet's climate, we show results for the Earth's case with $5 \times 10^{16}$ m$^3$ of water, which corresponds to a land fraction of 0.78 (the right column in Figure 10). This planet has a lower evaporation efficiency and cloud mass mixing ratio than that of an aqua planet, leading to the planet being able to have a larger amount of outgoing long-wave radiation. The boundary between an aqua planet's climate and a land planet's climate is thus located around a land fraction of 0.4. Even for the cases of Mars' and Venus' topographies, we recognized a land fraction of 0.4 as the boundary between the two climates (Figure 9).

If the condition of an aqua planet's climate would be a globally connecting ocean, the land fraction at the boundary between an aqua planet's climate and a land planet's climate should be 0.5 because half the area is needed to form a connecting region for a given space according to the percolation theory. However, we find that the climate state is determined by whether or not the land fraction exceeds 0.4 when the climate reaches the thermal equilibrium state, although the land fraction after a GCM calculation depends on the initial amount of water on the planetary surface.

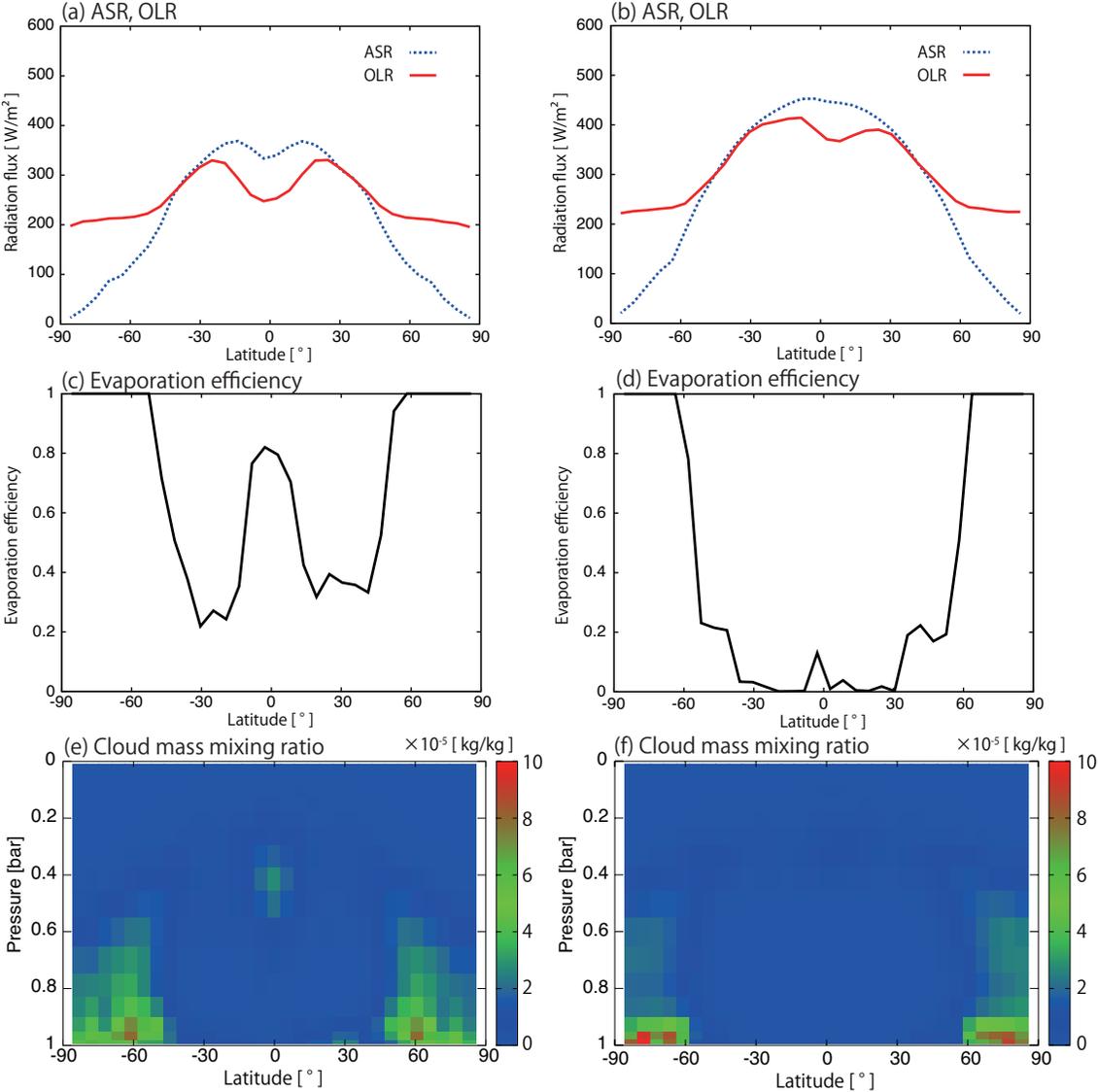

Figure 10. The differences of climate variables around the boundary between an aqua planet (left column) and a land planet (right column) for the Earth's topography. Each panel shows (a and b) the absorbed solar radiation (ASR) and the outgoing long-wave radiation (OLR); (c and d) the evaporation efficiency; (e and f) the cloud mass mixing ratio.

## 5 Discussion

### 5.1 Amount of water at the boundary of climates

Water planets are classified into three types: an ocean planet, a partial-ocean planet and a land planet. The surface of an ocean planet is globally covered with ocean, and a partial-ocean planet has a globally connected ocean with a continent. These two types of water planets are called aqua planets, and they have a common characteristic in possessing a globally connected ocean. *Kodama et al.* [2018] focused on the boundary between an aqua planet and a land planet that has a small amount of water on its surface; in other words, a globally disconnected ocean. They estimated the lowest latitude of the surface water distribution with different amounts of water pouring from the poles, which corresponds to the water flow limit, and investigated the relationship between the water flow limit and the amount of water on the planetary surface. They found that the boundary between two climates is controlled by the Hadley circulation. When the latitude of the water flow limit is located within the Hadley circulation, a terrestrial planet with such a surface water distribution behaves as an aqua planet. When the latitude of the water flow limit is located at the edge of the Hadley circulation, the amount of water on the planetary surface is ~1×10$^{16}$ m$^3$ (~1% of $V_{OCE}$) for the present Earth's topography. Thus, they estimated the amount of water at the boundary between an aqua planet and a land planet is ~1% of $V_{OCE}$.

*Kodama et al*. (2018) just poured water from the poles to estimate the relationship between the latitude of the water flow limit and the amount of water on the planetary surface for the topographies of the terrestrial planets. They did not perform the GCM calculations for 2-D water surface distribution. Here we calculated the atmospheric circulation after we determined the surface water distribution. Our results show that a water planet behaves as if it has a land planet's climate, even if the water flow limit reaches the equator. For the topography of the Earth, the water flow limit reaches the equator where a planet has 2×10$^{16}$ m$^3$ the amount of water on the planetary surface. However, a water planet with the Earth's topography and 5×10$^{16}$ m$^3$ the amount of water behaves as if it has the climate of a land planet at the runaway threshold because it has a wide dry area around the equator. On the other hand, a water planet with 1×10$^{17}$ m$^3$ the amount of water behaves as if it has the climate of an aqua planet at the runaway threshold.

For the topography of Mars, the boundary of climates is located in the amount of water on the planetary surface being between 5×10$^{17}$ m$^3$ and 1.37×10$^{18}$ m$^3$. For the topography of Venus, it is located in the amount of water between 5×10$^{16}$ m$^3$ and 1×10$^{17}$ m$^3$. The width of the boundary of the amount of water between an aqua planet and a land planet for Mars cases is a little larger than those for Earth and Venus cases because the topography of Mars has a higher altitude around the equator than those for the Earth and Venus. Thus, if we consider the topography of a terrestrial planet, the boundary of the amount of water between an aqua planet and a land planet should lie between about 1×10$^{17}$ m$^3$ and 1×10$^{18}$ m$^3$ of the amount of water on the planetary surface. A water planet with an amount of water less than 1×10$^{17}$ m$^3$ (~10% of $V_{OCE}$) should behave as if it has the climate of a land planet, even if it has a different distribution of water on its surface. Additionally, a water planet with an amount of water larger than 3 $V_{OCE}$ is globally covered with an ocean and the insolation at the runaway threshold is the value for an aqua planet, although we do not consider such a case in this study.

## 5.2 Effect of oceanic albedo on the runaway threshold

In all GCM simulations that we conducted in the previous sections, we set the surface albedo to be 0.3 except for the regions covered by snow, according to previous studies [*Abe et al*., 2011; *Kodama et al*., 2018]. However, the Earth's ocean has a lower albedo (~0.07) depending on the depth of the ocean.

We estimate the runaway threshold under a different setup from Section 4.2. For the pseudo-Earth setup, we set the surface albedo to be 0.07 in the water pool region and 0.3 on land except for the region coved by snow. In this setup a planet has Earth's topography, which includes altitude, and no vegetation. Hereafter, we call a set up in Section 4 the idealized-Earth setup to distinguish between them. Here, we discuss the effect of the oceanic albedo on the runaway threshold and the boundary between an aqua planet and a land planet.

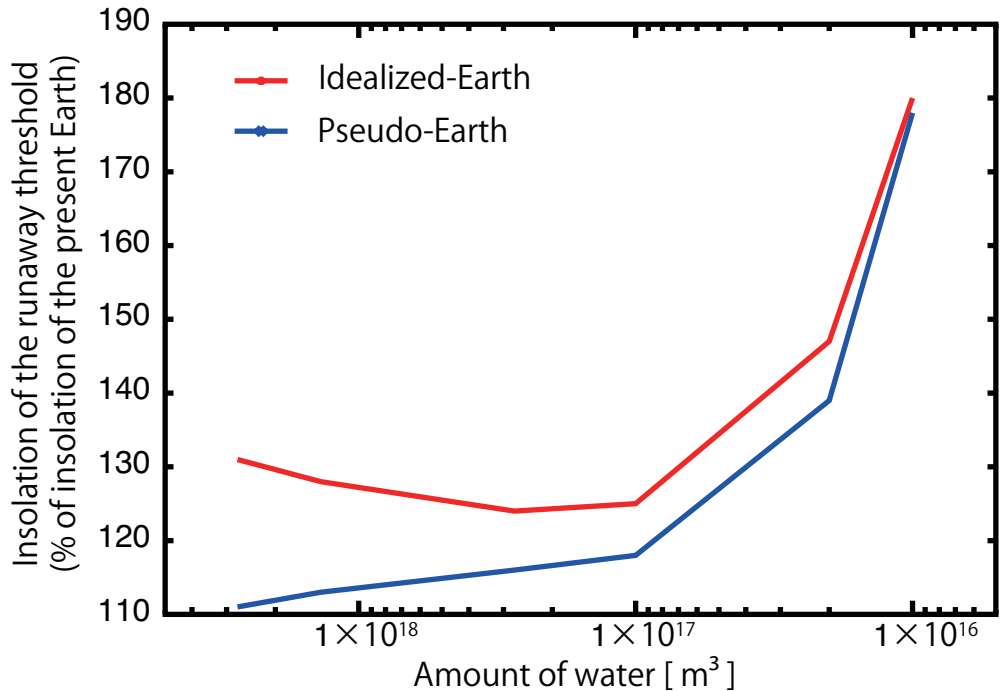

Figure 11. The runaway threshold for the idealized-Earth setup (red line) and the pseudo-Earth setup (blue line) with different amounts of water.

Figure 11 shows the runaway threshold under two setups. It shows a high value of the runaway threshold for both cases when the amount of water on the planetary surface is small. The runaway threshold for the pseudo-Earth experiment with the amount of water of 1 $V_{OCE}$ ($1.37 \times 10^{18}$ m³) is 113% $S_0$, which is a comparable value to the runaway threshold for previous studies that assumed the present Earth [*Leconte* et al., 2013b]. When the amount of water on the planetary surface decreases to 1% of $V_{OCE}$, the runaway threshold increases to 178% $S_0$. As shown in Figure 11, the runaway threshold strongly depends on the surface conditions.

When the amount of water is larger than 0.2 $V_{OCE}$, the runaway thresholds for the pseudo-Earth experiments decrease with the increase of the amount of water and, in contrast, those for the idealized-Earth experiments increase with an increasing amount of water. Figure 12 shows the surface albedo and the planetary albedo for both of the experiments with different amounts of water on the planetary surface. For example, the surface albedo and the planetary albedo for the pseudo-Earth experiment with 1 $V_{OCE}$ are 0.14 and 0.33, which are close to the global mean surface and planetary albedo on the present Earth. Figure 12 also shows that the increase in the amount of clouds with the increase in the amount of water for the idealized-Earth experiment causes the increase of the planetary albedo because of the scattering of the insolation from the central star and the increase of the runaway threshold. On the other hand, the surface albedo decreases with the increase of the amount of water for the pseudo-Earth experiment. This is because the sea surface albedo is lower than that for the desert. As a result, the runaway threshold decreases for the pseudo-Earth experiments. The difference between both of the experiments is smaller than the difference of the surface albedo between both of the experiments. It implies that the clouds reduce the effect of surface conditions on the runaway threshold.

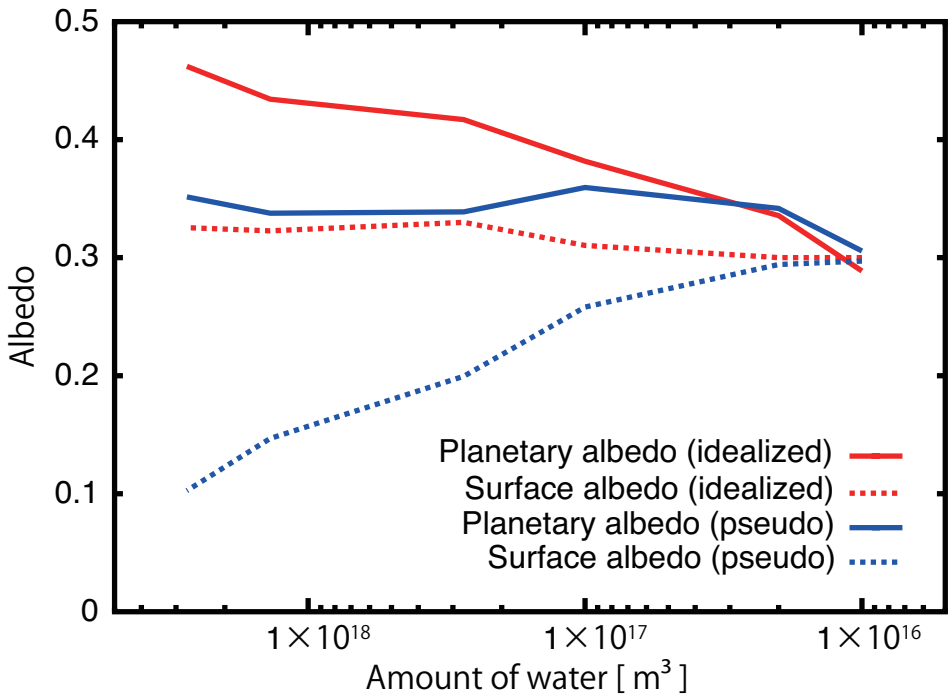

Figure 12. The planetary and surface albedos for the idealized-Earth setup and the pseudo-Earth setup for the different amounts of water.

### 5.3 Effect of recycling materials between the planetary surface and interior

In this study, we assumed an Earth-like atmosphere which includes fixed 1 bar of $N_2$ and no $CO_2$ for all simulations. However, the variation of the concentration of atmospheric $CO_2$ ($pCO_2$) is also important for the planetary habitability. $pCO_2$ would be higher on aqua planets than on planets with oceans and land [e.g., *Tajika and Matsui*, 1992] because the silicate weathering is faster on continents than on the seafloor. However, on a planet with a high land fraction, it is not clear whether $pCO_2$ in the atmosphere is higher or lower because it depends on the activity of recycling processes of materials between the planetary surface and interior. When the activity of this recycling is similar to the current Earth, planets with high land fractions have higher $pCO_2$ levels than that of planets with oceans and land because there would not be enough rainfall to cause weathering on the planetary surface. On the other hand, when it is weaker than that for the current Earth because of lack of water in the mantle, the concentration of $CO_2$ in the atmosphere is hard to estimate because the amount of rainfall decreases, but the outgassing of water vapor also decreases. According to *Kasting* [1988], $pCO_2$ has little effect on the runaway greenhouse threshold. Thus, the concentration of atmospheric $CO_2$ does not affect results from our simulations. However, it would have an effect on the moist greenhouse limit [*Popp et al.*, 2016]. $pN_2$ also affects this limit as a background atmosphere. Therefore, we have to consider these components to investigate the long-term evolution of habitable planets.

Additionally, the stability of very small ocean still remains an issue because a small amount of water would be trapped in silicate minerals as hydrous minerals, which are subducted into the planetary interior. If plate tectonics work well on a land planet, a small amount of water would be subducted into the planetary interior, but it is hard to say whether such a small ocean is still stable or not because there is an outgassing of water from the planetary interior to the atmosphere. If the planetary interior is already wet, small oceans can also survive. If the plate tectonics do not work well, the hydrated region is limited to the planetary surface, which would not be able to hydrate as much as 10% of the Earth's ocean mass on the land planet.

## 5.4 Implications for exoplanets

A number of exoplanets have been detected and some of them are thought to be Earth-sized or super Earth-sized rocky planets. Recently, a few tens of Earth-sized and super-Earth-sized planets orbiting in the classical habitable zone for an aqua planet have been reported. For Proxima Centauri b and terrestrial planets in the TRAPPIST-1 system, their climates have been investigated by GCM, and their habitability has also been discussed [*Turbet et al.*, 2016; 2018]. In the next decade, the number of detections of exoplanets should increase via observation by TESS (Transiting Exoplanets Survey Satellite) and PLATO (PLAnetary Transits and Oscillations of stars) and the characterization of exoplanets will advance thanks to the JWST (James Webb Space Telescope) and TMT (Thirty Meter Telescope).

A land planet has a wider habitable zone than that of an aqua planet, which is a strong advantage for characterizing habitable planets and detecting biomarkers as the number of potential habitable planet increases. Our results show that the inner edge of the habitable zone is strongly affected by the surface water distribution. The pseudo-Earth experiments with a large amount of water and the previous studies for the runaway threshold of the present Earth estimate the runaway threshold to be about 110% $S_0$, which corresponds to about 0.95 AU when the age of a G type star is 4.5 Gyrs. The idealized-Earth experiments with a large amount of water show the runaway threshold to be about 130% $S_0$, which corresponds to about 0.88 AU. The runaway threshold increases from about 130% $S_0$ to 180% $S_0$ with decreasing surface water area and the inner edge of the habitable zone extends to about 0.75 [AU]. The luminosity of a Sun-like star increases with time. The luminosity of a Sun-like star at 10 Myr—which is thought to be the timing of the disappearance of the protoplanetary disk—is about 976 W/m², which is darker than that of the present Sun. At 10 Myr, the inner edges of the habitable zone for each state are 0.80 AU, 0.74 AU and 0.63 AU for about 110%, 130% and 180% $S_0$, respectively. This suggests that a planet with a small amount of water can maintain liquid water on its surface even if it is located at Venus' orbit.

*Kodama et al.* [2015] investigated the evolution from an aqua planet to a land planet via rapid water loss. They assumed the typical amount of water on the surface of a land planet to be 5% of $V_{OCE}$ based on a result from *Abe et al.* [2011]. They found that an aqua planet with an initial amount of water below 1% $V_{OCE}$ can evolve into a land planet and maintain liquid water on its surface for another 2 Gyrs. Our experiments showed that a water planet with below about 10% $V_{OCE}$ has the climate of a land planet. Therefore, the maximum initial amount of water to evolve from an aqua planet to a land planet should become larger, and extended area of the inner edge of the habitable zone due to the evolution from an aqua planet to a land planet is also wider than that estimated by *Kodama et al*. [2015].

There are some important factors to determine the climate of a water planet, such as the planetary rotation rate, the planetary mass, the surface gravity, the mass of the planetary atmosphere, obliquity and so on. Although we can estimate the rough bulk composition of an exoplanet by the relationship between the planetary mass and radius, it is difficult to distinguish such a small amount of water on the planetary surface compared to the planetary mass (even for 1 $V_{OCE}$). We showed that the amount of water on a planet has a strong influence on the climate of a water planet. The range of the amount of water that we consider is between an order of $10^{-4}$ and $10^{-2}$ wt% of the planetary mass. Thus, a small difference in the amount of water forms a significant difference in the climate of a water planet and we can give a strict constraint of the amount of water, which is hard to detect, by using the large difference in climate between an aqua planet and a land planet. Our study suggests that the large differences between both of the climates are due to the distribution of clouds, ocean and land. A strategy to detect the distribution of ocean and land has been proposed and it can distinguish between an aqua planet and a land planet [*Fujii* et al., 2010]. To investigate the character of a land planet through observation should be a future task.

## 6 Summary

Previous studies [*Abe et al*., 2011; *Kodama et al*., 2018] showed the climate of a planet with a small amount of water on its surface, which is called a land planet, is significantly different from that of an aqua planet. Because a land planet can emit a large amount of radiation from the dry tropics exceeding the Simpson-Nakajima limit which is a limit of radiation from a wet atmosphere, it has a larger runaway threshold than that of an aqua planet. This means that the inner edge of the habitable zone for a land planet is located closer to the central star than that of an aqua planet.

However, since they only investigated a few limited cases for the surface water distribution, the relationship between the runaway threshold and the surface water distribution is still not clear. Here, we considered two different types of surface water distribution: one is the meridionally uniform surface water distribution and the other is a water distribution that is determined by the topographies of terrestrial planets in our solar system. We systematically investigated the runaway threshold for these cases using a GCM.

For the meridionally uniform surface water distribution, we considered two extreme cases for the water distribution on the surface: the equally dispersed case and the concentrated case. Even if they have the same area of wet regions, the runaway threshold for the dispersed case is lower than that for the concentrated case because the former has a large amount of water vapor in the tropics that is sufficient to limit the planetary radiation. The runaway threshold for the equally dispersed case increases from about 130% $S_0$, which corresponds with that of an aqua planet, to about 155% $S_0$ (the land-planet regime). Although we recognized two climate regimes, the obtained runaway threshold for the land-planet regime is quite a lot lower than that of the previous works [*Kodama et al*., 2018] because the atmosphere above part of the tropics is always wet for a meridionally uniform case. We found that the runaway threshold for a meridionally uniform dispersed case corresponds with the lower limit of the runaway threshold depending on the surface water distribution because this surface distribution efficiently wets the atmosphere. As *Kodama et al*. [2018] showed, the runaway threshold for a zonally uniform surface water distribution corresponds with the upper limit of the runaway threshold because of the dry tropics. Therefore, we showed a range of possible runaway thresholds depending on the surface water distribution.

We found that the runaway threshold for the surface water distribution determined by Earth's, Mars' and Venus' topographies is close to that for the meridionally uniform cases because it tends to create a wet region near the tropics. We found that a water planet with a land fraction less than 0.4 behaves as an aqua planet, and a water planet with a land fraction more than 0.4 behaves as a land planet, where the land fraction is the ratio of the dry surface area over the total area. Depending on the topography, the amount of water at the boundary between an aqua planet and a land planet is typically around 10% the amount of the Earth's ocean.

Our results show that a small difference in the amount of water on the planetary surface forms a significant difference in the characteristics of climates. Once future observations can distinguish this difference in climate characteristics, this will be a strong constraint on the amount of surface water, which is hard to estimate from the relationship between the planetary radius and mass.

## Acknowledgments, Samples, and Data

We thank Prof. James Kasting and an anonymous reviewer for their thorough review and constructive comments. This work was supported by MEXT KAKENHI grant JP17H06104 and JP17H06457. This project has received funding from the European Research Council (ERC) under the European Union's Horizon 2020 research and innovation programme (grant agreement No.679030/WHIPLASH). A GCM used in this paper is CCSR/NIES AGCM 5.4g which has been developed for the Earth's climate by the Center for Climate System Research, the University of Tokyo, and the National Institute for Environmental Research. We used Gtool3-dcl5 which is developed by the GFD-DENNOU Club for analysis (https://www.gfd-dennou.org). Data sets to create typical figures in this study are available on website (http://doi.org/10.5281/zenodo.3326630).

## References

Abbot, S. D., N. B. Cowan, and F. J. Ciesla (2012), Indication of insensitivity of planetary weathering behavior and habitable zone to surface land fraction, *Astrophys. J.*, 756:178.

Abe-Ouchi, A., F. Saito, K. Kawamura, M. E. Raymo, J. Okuno, K. Takahashi, and H. Blatter (2013), Insolation-driven 100,000-year glacial cycles and hysteresis of ice-sheet volume, *Natute*, 500(7461), 190-193.

Abe, Y., and M. Matsui (1988), Evolution of an impact-generated $H_2O$-$CO_2$ atmosphere and formation of a hot proto-ocean on Earth, *J. Atmos. Sci.*, 45, 3081-3101.

Abe, Y., A. Numaguti, G. Komatsu, and Y, Kobayashi (2005), Four climate regimes on a land planet wet surface: effect of obliquity change and implications for Mars, *Icarus*, 178, 27-39.

Abe, Y., A. Abe-Ouchi, N. H. Sleep, and K. J. Zahnle (2011), Habitable zone limits for dry planets, *Astrobiology*, 11(5): 443-460.

Arakawa, A., and W. H. Schubert (1974), Interaction of cumulus cloud ensemble with the large-scale environment, part I, *J. Atmos. Sci.*, 31, 674-701.

Batalha, N. M., et al. (2013), Planetary candidates observed by Kepler. III. Analysis of the first 16 months of data, *Astrophys. J.* 204:24.

Fujii, Y., H. Kawahara, Y. Suto, A. Taruya, T. Nakajima, and E. L. Turner (2010), Colors of a second Earth: Estimating the fractional areas of ocean, land, and vegetation of Earth-like exoplanets, *Astrophys. J.*, 715, 866-880.

Goldblatt, C., T. D. Robinson, K. J. Zahnle, and D. Crisp (2013), Low simulated radiation limit for runaway greenhouse climates, *Nature Geoscience*, 6: 661-667.

Haltiner, G. J., and R. T. Williams (1980), Numerical Prediction and Dynamic Meterology, *Jhon & Sons*: 477pp.

Hunten, D. M. (1973), The escape of light gases from planetary atmospheres, *J. Atmos. Sci.*, 30, 1481-1494.

Hirt, C., M. Kuhn, W. E. Featherstone, and F. Göttl (2012), Topographic/isostatic evaluation of new-generation GOCE gravity field models, *J. Geophys. Res.*, 117, B05407.

Ingersoll, A. P. (1969), The runaway greenhouse: a history of water on Venus, *J. Atmos. Sci.*, 26: 1191-1198.

Kasting, J. F. (1988), Runaway and moist greenhouse atmospheres and the evolution of Earth and Venus, *Icarus*, 74: 472-494.

Kasting, J. F., D. P. Whitmire, and R. T. Reynolds (1993), Habitable zones around main sequence stars, *Icarus*, 101, 108-128.

Kasting, J. F., H. Chen, and R. K. Kopparapu (2015), Stratospheric temperature and water loss from moist greenhouse atmospheres of Earth-like planets, *Astrophys, J. Lett.*, 813:L3.

Kodama, T., H. Genda, Y. Abe, and K. J. Zahnle (2015), Rapid water loss can extend the lifetime of planetary habitability, *Astrophys. J.,* 812:165.

Kodama, T., A. Nitta, H. Genda, Y. Takao, R. O'ishi, A. Abe-Ouchi, and Y. Abe (2018), Dependence of the onset of the runaway greenhouse effect on the latitudinal surface water distribution of Earth-like planets, *J. Geophys. Res. Planets*, 123, 559-574.

Komabayashi, M. (1967), Discrete equilibrium temperature of a hypothetical planet with the atmosphere and hydrosphere of a one component-two phase system under constant solar radiation, *J. Meteor. Soc. Japan*, 45, 137-139.

Kopparapu, R. K., R. Ramirez, J. F. Kasting, V. Eymet, T. D. Robinson, S. Mahadevan, R. C. Terrien, S. Domagal-Goldman, V. Meadows, and R. Deshpande (2013), Habitable zones around main-sequence stars: New estimates, *Astrophys, J.*, 765, 2.

Kopparapu, R. K., R. M. Ramirez, J. S. Kotte, J. F. Kasting, S. Domagal-Goldman, and V. Eymet (2014), Habitable zones around main-sequence stars: Dependence on planetary mass, *Astrophys, J. Lett.*, 787:L29.

Leconte, J., F. Forget, B. Charnay, R. Wordsworth, F. Selsis, E. Millour, and A. Spiga (2013a), 3D climate modeling of close-in land planets: Circulation patterns, climate moist bistability, and habitability, *A&A*, 554, A69.

Leconte, J., F. Forget, B. Charnay, R. Wordsworth, and A. Pottier (2013b), Increased insolation threshold for runaway greenhouse processes on Earth-like planets, *Nature*, 504, 268-271.

Le Treut, H., and Z.-X. Li (1991), Sensitivity of an atmospheric general circulation model to prescribed SST changes: feedback effects associated with the simulation of cloud optical properties, *Clim. Dyn.*, 5: 175-187.

Manabe, S. (1969), Climate and the ocean circulation. I. The atmospheric circulation and the hydrology of the Earth's surface, *Mon. Wea. Rev.*, 97, 739-774.

Moorthi, S., and M. J. Suarez (1992), Relaxed Arakawa-Schubert: A parameterization of moist convection for general circulation models, *Mon. Wea. Rev.*, 120, 978-1002.

Nakajima, T., and M. Tanaka (1986), Matrix formulations for the transfer of solar radiation in a plane-parallel scattering atmosphere, J. Quant. Spectros. Radiant. Transfer, 35, 1, 13-21.

Nakajima, S., Y.-Y. Hayashi, and Y. Abe (1992), A study on the runaway greenhouse effect with one-dimensional radiative-convective equilibrium model, *J. Atmos. Sci.*, 49(23), 2256-2266.

Numaguti, A. (1999), Origin and recycling processes of precipitating water over the Eurasian continent: Experiments using an atmospheric general circulation model, *J. Geophys. Res.*, 104, D2, 1957-1972.

Popp, M., H. Schmidt, and J. Marotzke (2016), Transition to a Moist Greenhouse with CO2 and solar forcing, *Nature Communications*, 7, 10627.

Tajika, E., and T. Matsui (1992), Evolution of terrestrial proto-$CO_2$ atmosphere coupled with thermal history of the earth, *EPSL*, 113, 251-266.

Turbet, M., J. Leconte, F. Selsis, E. Bolmont, F. Forget, I. Ribas, S. N. Raymond, and G. Anglada-Escudé (2016), The habitability of Proxima Centauri b. II. Possible climates and observability, *A&A*, 596, A112.

Turbet, M., E. Bolmont, J. Leconte, F. Forget, F. Selsis, G. Tobie, A. Caldas, J. Naar, and M. Gillon (2018), Modeling climate diversity, tidal dynamic and the fate of volatiles on TRAPPIST-1 planets, *A&A*, 612, A86.

Walker, J. C. G. (1977), Evolution of the Atmosphere, MacMillan, New York.

Way, M. J., A. D. Del Genio, I. Aleinov, T. L. Clune, M. Kelley, and N. Y. Kiang (2018), Climates of Warm Earth-like Planets. I. 3D Model Simulations, *Astrophys, J. SS.* 239, 24.

Wieczorek, M. A. (2007), The gravity and topography of the terrestrial planets, *Treatise on Geophys.*, 10: 165-206.

Wolf, E. T., and O. B. Toon (2014), Delayed onset of runaway and moist greenhouse climates for Earth, *Geophys. Res. Lett.*, 41, 167-172.

Wolf, E. T., and O. B. Toon (2015), The evolution of habitable climates under the brightening Sun, *J. Geophys. Res. Atmos.*, 120, 5775-5794.